\documentclass[twocolumn]{revtex4}
\usepackage{amsthm}
\usepackage{amsmath}
\usepackage{rotating}
\usepackage{natbib}
\usepackage{setspace}

\begin{document}

\title{Overcoming Problems in the Measurement of Biological Complexity}

\author{Manuel Cebrian}
\author{Manuel Alfonseca}
\author{Alfonso Ortega}
\affiliation{%
Department of Computer Science, Universidad
  Aut\'onoma de Madrid, 28033 Madrid, Spain}

\begin{abstract}
In a genetic algorithm, fluctuations of the entropy of a genome over
time are interpreted as fluctuations of the information that the
genome's organism is storing about its environment, being this
reflected in more complex organisms. The computation of this entropy
presents technical problems due to the small population sizes used in
practice. In this work we propose and test an alternative way of
measuring the entropy variation in a population by means of
algorithmic information theory, where the entropy variation between
two generational steps is the Kolmogorov complexity of the first step
conditioned to the second one. As an example application of this technique, we report experimental differences
in entropy evolution between systems in which sexual reproduction is
present or absent.
\end{abstract}

\maketitle

\section{Introduction}

The evolution over time of the entropy of a genome within a population is currently 
an interesting problem which is conjectured to be connected to
the evolution of the complexity of organisms in a genetic algorithm
\citep{ADAMI:OFRIA,ADAMI}. The complexity of the genome of an organism
is considered to be the amount of
information about its environment it stores. That is, evolution would cause the
appearance of more complex sequences, which correspond to more complex
phenotypes. This hypothesis states that natural selection acts as a Maxwell
demon, accepting only those changes which adapt better to the environment and
give rise to more complex individuals with genomes of lower entropy.
This idea was tested by the simulation of a very simple system of
asexual individuals in fixed environmental conditions.

However, it is well known that that the computation of the entropy as
\begin{equation}
H(\text{genome})=-\sum_{x \in \text{genomes in the population}} p(x) \log
p(x)
\label{complete}
\end{equation}
has technical complications, due to the large size of the sample
needed to estimate it with accuracy \citep{ADAMI,herzel1994ebr,basharin1959see}.
In practice, it is usually estimated as 

\begin{equation}
H(\text{genome})=\sum_{i=1..\text{size(genome)}} H(i),
\label{partial}
\end{equation}
i.e., as the sum of the entropy
contributions of each locus in the genome. This estimation misses the
entropy contributions due to epistatic effects. Some sophisticated
statistical methods can be used to remedy this (see Appendix in
\cite{ADAMI:OFRIA}), although we will not
deal with them in this work. 

An still unexplored way to overcome this problem is to estimate the entropy of a
genome as its average Kolmogorov complexity 
\begin{equation}
\left <K(\text{genome})\right>_{\text{genomes in the population}} \rightarrow
 H(\text{genome})
\label{kolm}
\end{equation} 
(see \cite{cover1991eit,kolmogorov1968taq,li1997ikc}). 
However, this result only holds for infinitely long sequences,
and therefore it cannot be applied to finite (sometimes short) genomes.

If we are only interested in the entropy evolution of the genome, and
not in the particular value estimation, we can resort to the
following trick: the genetic algorithm can be modelled as a thermodynamic system
which evolves over time, where every population is just a measurement by
an observer, i.e. the system is modeled as a statistical
ensemble of genomes and each measurement is just a sample from that
ensemble.

Now we can measure the entropy evolution of the system from two different
viewpoints. The first is the system itself, where
the entropy is calculated a la Shannon (equation \ref{partial}) by
estimating the probabilities of the loci alleles, using the
frequencies of the ensemble sample.

The second way of measuring the entropy is from the viewpoint of the
observer, where a measurement is made of the
population at each time step, and the information about the
system is updated, i.e., the observer measures the system at time $t$ and stores
this information $S_t$. At time $t+1$ the observer makes another
measurement and substitutes $S_t$ by $S_{t+1}$. The entropy variation
due to this substitution can be calculated for both equilibrium and
non-equilibrium thermodynamic systems \citep{ZUREK1,ZUREK3,bennett1982tcr}.
Since evolution cannot be modeled as a
system in equilibrium, the second case applies: the mutation and generational replacement 
operators may increase or decrease the entropy of the system
\citep{vose1999sga,wright2005fga}.

Thus the entropy variation from the
observer viewpoint is $K(S_t)-K(S_{t+1})$ bits. As $K(\cdot)$ is
an incomputable measure, we estimate it by using the Lempel-Ziv
algorithm \citep{ZIV} as $\lim_{n \rightarrow \infty}
(1/n)|LZ(x)| = (1/n) K(x|n)$, where $x$ is an infinite string and
$|LZ(x)|$ is
the size of the same string compressed \citep{cover1991eit}. Now our
measurement of the system at time $t$, $|S_t|$ is much larger than with
equation \ref{kolm} (just the genome), so the estimation becomes possible.

\section{Experiments}

We want to test whether the evolution of $K(\text{population sample})$ can help in
the study of the evolution of $H(\text{genome})$. Both measurements
have their limitations,
but their agreement would provide evidence that the entropy
evolution is being studied correctly.

We have evaluated this experimentally, using the 
genetic algorithm proposed by \cite{HAYASHI:VOSE}, 
which is able to reproduce sexual behaviour in a 
vary detailed way, because it includes several features absent 
in the \cite{ADAMI:OFRIA} experiments, such as sexual reproduction, different 
inter-locus and intra-locus interactions across the genotypic
or phenotipyc distance, and the evolutionary mechanisms of 
mutation and natural selection.

\begin{figure}[t]
\centering
\includegraphics[width=0.5\textwidth]{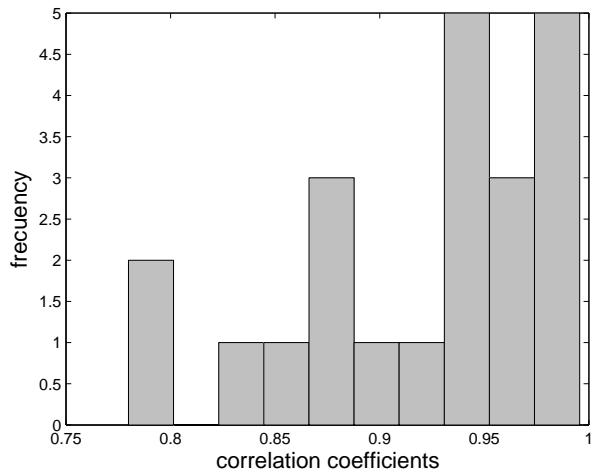}
\caption{Histogram of the 22 experimental correlation coefficients.}
\label{corrcoef}
\end{figure} 

We have implemented and run the same simulation proposed by 
Hayashi et al. The model's sexual dynamics 
can be summarized as follows: there are two different sexes (male and
female). The likelihood that a female with trait $a$ will mate with a male with
trait $b$ is defined by $\psi(d)=e^{-\alpha d^{2}}$, where $d$ is the genetic
or phenotypic distance measuring compatibility between the sexes, and 
$\alpha$ is a parameter that represents the compatibility between traits.
The value of $\alpha$ used in the simulations is $0.005$. The overall 
number of offspring produced by a female in each coupling is given by
$W_{f}=B_{max}e^{-sc(P-P_{opt})^{2}}$, where $P$ is the proportion of males with
which the female has been able to mate. The parameter $P_{opt}$ (which can take the values 
$\{0.2, 0.4, 0.6, 0.8\}$) defines the 
fertility of the female. The parameter $sc$ (ranging from $1.02$ to $4 \times 1.02$) 
stands for sexual conflict selection
in females. $B_{max}$ is the maximum possible number of offspring ($5$).
The sex and the father are randomly chosen for each offspring, and the 
number of males each female encounters is $n=20$. 

Twenty two different experiments have been performed. $K(\text{population sample})$ 
and $H(\text{genome})$ have been estimated for each generation 
step with the methods described above. 
The system evolved for 10,000 generations
with a population size of 1,000 individual genomes.

In all our results, both measures were highly correlated (see
fig. \ref{corrcoef}), giving 
evidence that the dual measurement of the evolution of entropy by
means of Kolmogorov complexity confirms the use of the Shannon
entropy.

\begin{figure}[t]
\centering
\includegraphics[width=0.5\textwidth]{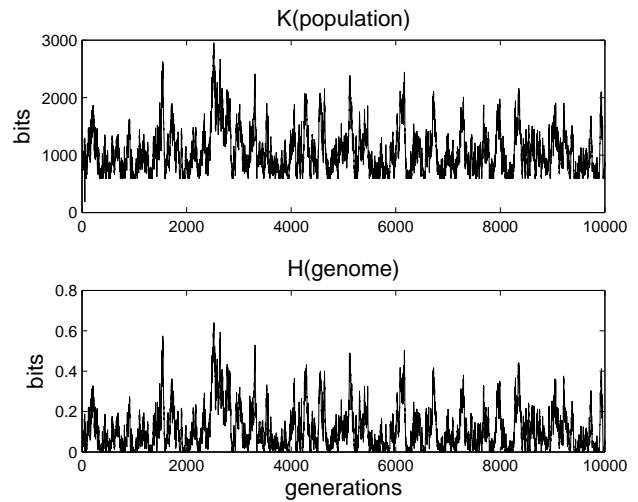}
\caption{Example of \emph{continuous evolutionary chase} without
  \emph{genetic differentiation} obtained with
parameters $B_{max}=5$, $P_{opt}=0.2$, $\mu=5.0 \times 10^{-5}$,
$\alpha=0.05$, $sc=1.02$, 2 loci, phenotypic model: additive. Experimental
correlation coefficient: 0.9956.}
\label{additive}
\end{figure}

\section{Discussion}

\begin{figure}[t]
\centering
\includegraphics[height=7cm]{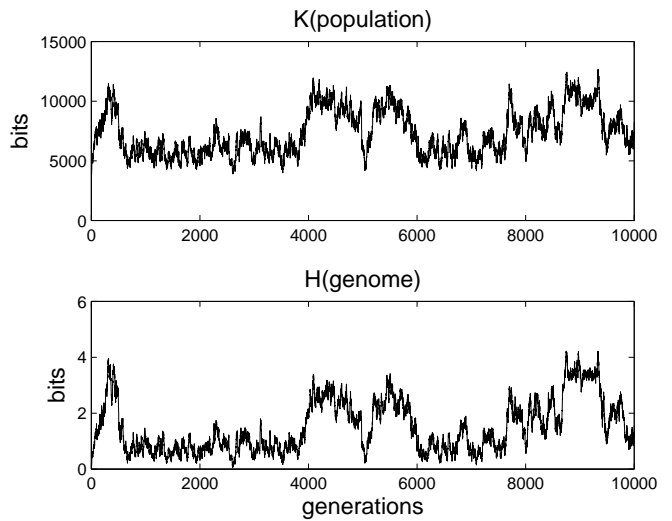}
\caption{Example of \emph{differentiation}  without
  \emph{speciation}, obtained with
parameters $B_{max}=5$, $P_{opt}=0.8$, $\mu=5.0 \times 10^{-5}$,
$\alpha=0.05$, $sc=1.02$, 8 loci, phenotypic model: co-dominance. Experimental
correlation coefficient: 0.9741.}
\label{codominance}
\end{figure}

When sexual dynamics are introduced in the system, the increase of 
complexity observed by
\cite{ADAMI:OFRIA} is not present anymore.

\begin{figure}[t]
\centering
\includegraphics[width=0.5\textwidth]{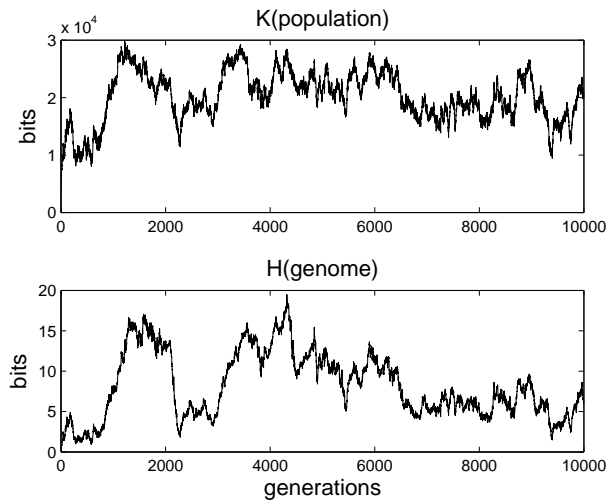}
\caption{Example of \emph{genetic differentiation} without \emph{co-evolutionary chase} or \emph{simpatryc
speciation}, obtained with parameters $B_{max}=5$, $P_{opt}=0.2$,
$\mu=5.0 \times 10^{-5}$, $\alpha=0.05$, $sc=1.02$, 32 loci,
phenotypic model:
dominance. Experimental correlation coefficient: 0.8571.}
\label{dominance}
\end{figure}

The typical result observed in our experiments is a chaotic behavior
of the entropy (fig. \ref{additive}). Only large genome lengths or
high female mating rates ($P_{opt}$) escape from this (the other
parameters seem to have little importance). The effect of this is to
increase the autocorrelation of the entropy time series and
provoke the rise of a few entropy \emph{bumps} during a small
number of generations (figs. \ref{codominance} and \ref{dominance}).

\begin{figure}[t]
\centering \includegraphics[width=0.5\textwidth]{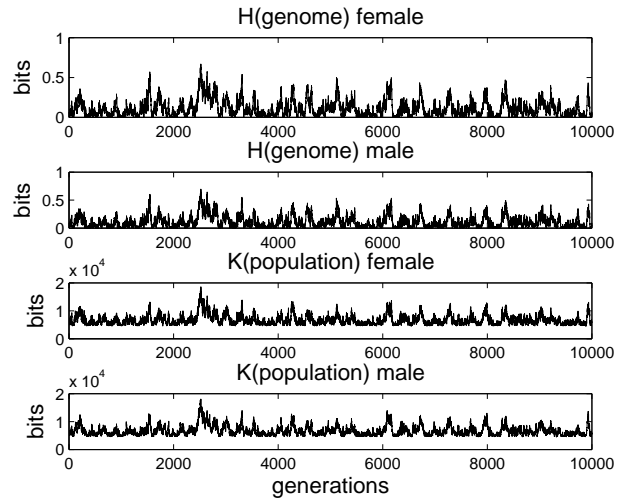}
\caption{Same parameters as in fig. 2 with entropy calculation decomposed by sex.}
\label{decomposed}
\end{figure}

Our hypothesis for this behavior of the entropy is the absence of
natural selection in the \citet{HAYASHI:VOSE} model, which could
explain the similarities between male and female evolution
(fig. \ref{decomposed}). Without natural selection, the environment for
females is reduced to random boundary conditions (mutations). On the
other hand, males are selected by females as mating partners. In this
way, females can be considered to become the environment for males,
since they determine the way in which the entropy of the males evolves. On
the other hand, females have no environment to adapt to. Perhaps if
the pressure of natural selection was applied both to male and female
(not necessarily in the same way), more complex patterns in the
behavior of their entropies would appear. The fact that 
natural selection is not taken into account may be the cause 
of the differences in entropy evolution between the models by 
\cite{ADAMI:OFRIA} and \cite{HAYASHI:VOSE}, and
the reason why global decreases in entropy are not observed in the latter.

\section{Conclussions and future work}

Studying a genetic algorithm from the observer point of view allows us to have large-scale estimates of the entropy evolution via Kolmogorov Complexity. This overcomes many limitations that arise from epistatic effects between loci and provide an easy way to do study the dynamics without resorting to complex mathematical trickeries. 
We also use this methodology to study what is the effect of sexual reproduction in terms of the evolution of complexity, as a decrease of entropy. We show that when sexual reproduction is present the population enters in a chaotic regime of complexity driven by the complexity drifts of the female organisms. We suggest that this might be cause for female organism evolving chaotically without natural selection, and male organisms evolving with females as the boundary conditions, which gives an overall chaotic evolution of complexity.
The next immediate step is to introduce natural selection in the experiments and study whether this will change the evolution of complexity. We plan to do it by implementing the typical natural selection operators from genetic algorithms such as tournament selection, steady state-selection and so forth. We conjecture that introducing natural selection will remove the chaotic complexity dynamics and might probably get closer to what Hayasi et. al. formerly reported: an increase in complexity by removing genetic mutations that do not improve the fitness function.

\section*{Acknowledgements}
We would like to thank Carlos Casta\~neda for his help in the
implementation and simulations.

\bibliography{main}
\bibliographystyle{plainnat}
\end{document}